\documentclass[twocolumn,prl,aps,showpacs,floatfix,preprintnumbers]{revtex4} 
\begin{document} 

%%%%%%%%%%%%%%%%%%%%%%%%%%%%%%%%%%%%%%%%%%%%%%%%%%%%%%%%%%%%
\title{Nonlocal pseudopotentials and magnetic fields}
%%%%%%%%%%%%%%%%%%%%%%%%%%%%%%%%%%%%%%%%%%%%%%%%%%%%%%%%%%%%

\author{Chris J. Pickard} 
\affiliation{TCM Group, Cavendish Laboratory, Madingley Road,\\
Cambridge, CB3 0HE, United Kingdom}

\author{Francesco Mauri} 
\affiliation{Laboratoire de Min\'{e}ralogie-Cristallographie de Paris,\\ 
Universit\'{e} Pierre et Marie Curie, 4 Place Jussieu, 75252, Paris, 
Cedex 05, France}

\date{\today}

\begin{abstract}
We show how to describe the coupling of electrons to non-uniform magnetic 
fields in the framework of the widely used norm-conserving pseudopotential approximation for electronic structure calculations. 
Our derivation applies to magnetic fields that are smooth on the
scale of the core region. 
The method is validated by application
to the calculation of the magnetic susceptibility of 
molecules. Our results are compared with high
quality all electron quantum chemical results, and another recently
proposed formalism.
\end{abstract}
\pacs{71.15.-m, 71.45.Gm, 71.15.Dx, 71.15.Mb} 
\maketitle

% 71.15.-m Methods of electronic structure calculations
% 71.15.Dx Computational methodology 
% 71.15.Mb Density functional theory
% 75.  Magnetic properties and materials
% 75.20.-g Diamagnetism, paramagnetism, and superparamagnetism 
% 71.45.Gm Exchange, correlation, dielectric and magnetic response functions, plasmon

%%%%%%%%%%%%%%%%%%%%%%%%%%%%%%%%%%%%%%%%%%%%%%%%%%%
\newcommand{\ket}    [1]{{|#1\rangle}}            % 
\newcommand{\bra}    [1]{{\langle#1|}}            %  
\newcommand{\braket} [2]{{\langle#1|#2\rangle}}   %
\newcommand{\bracket}[3]{{\langle#1|#2|#3\rangle}}%
%%%%%%%%%%%%%%%%%%%%%%%%%%%%%%%%%%%%%%%%%%%%%%%%%%%

%%%%%%%%%%%%%%%%%%%%%%%%%
%\section*{Introduction}%
%%%%%%%%%%%%%%%%%%%%%%%%%

The coupling of the electrons in matter with probing electromagnetic
fields or charged particles provides the basis for nearly all known
analytical experimental techniques. However, the most computationally
efficient schemes for the first principles prediction (and hence
interpretation) of the experimental observables require the use of
approximate, and crucially, nonlocal Hamiltonians. 
Most notably, the use of nonlocal pseudopotentials, along with with density functional
theory, is often referred to as the Standard Model of modern
electronic structure theory. This is not without justification. The
computational efficiency, and accuracy of modern first principles
pseudopotentials has allowed a vast range of problems to be solved for
realistic materials. However, whatever the successes of this method,
it is still unclear how nonlocal Hamiltonians should be coupled to
arbitrary magnetic fields.
In the specific case of  uniform magnetic fields described with 
the symmetric-gauge vector-potential, we have derived the correct Hamiltonian by developing
what we called the gauge including projector augmented wave (GIPAW) method~\cite{pickard2001-1}.
In this Letter we derive the pseudopotential Hamiltonian that
describes the coupling between the electrons and a 
 smooth non-uniform magnetic field represented by an arbitrary 
vector-potential gauge.

We are not alone in this quest. Recently Ismail-Beigi, Chang and
Louie~\cite{ismail-beigi2001}, to whom we refer the reader for a more
complete summary of earlier work in this field, proposed a scheme
(hereafter referred to as the ICL method) which sought to finally
provide a rigorous derivation of a closed form for the coupling of
nonlocal systems to arbitrary electromagnetic fields. While the
widespread use of nonlocal pseudopotentials clearly provided the
motivation and applications for their method, they attempted to tackle
the problem of a general nonlocal Hamiltonian. In contrast, this
Letter focuses exclusively on the class of nonlocal Hamiltonians that
arise due to the use of first-principles nonlocal pseudopotentials
for electronic structure calculations.

%%%%%%%%%%%%%%%%%%%%%%%%%%%%%%%%%%%%
%\section*{Generalised GIPAW proof}%
%%%%%%%%%%%%%%%%%%%%%%%%%%%%%%%%%%%%

In the pseudopotential approach, and in the absence of a magnetic field,
the all-electron (AE) Hamiltonian $H^{\rm AE}={p}^2/2 + V({\bf
r})$, is replaced by its pseudo (PS) equivalent, 
$ H^{\rm PS}={
p}^2/2 + V^{\rm l}({\bf r})+\sum_{\bf R} V^{\rm nl}_{\bf R}$.
$V^{\rm l}({\bf r})$ is a local-potential, and
$V^{\rm nl}_{\bf R}$ is a nonlocal operator which
acts only within the core region of the atomic sites, at ${\bf R}$: 
\begin{equation}
V^{\rm nl}_{\bf R}=
\int d^3r'd^3r'' |{\bf r}'\rangle \langle {\bf r}''| V^{\rm nl}_{\bf R}({\bf
r}',{\bf r}''). 
\end{equation}
By construction, in the valence-energy-range and to within some
controllable error, (i) the eigenvalues of $H^{\rm PS}$ coincide
with those of $H^{\rm AE}$, and (ii) the eigenstates of $H^{\rm PS}$
coincide with those of $H^{\rm AE}$ outside the core regions.

Turning on a magnetic field, ${\bf B}({\bf r})=\nabla \times {\bf
A}({\bf r})$, the AE Hamiltonian becomes:
\begin{equation}
H^{\rm AE}_{\bf A} = \frac{1}{2}\left[{\bf p}+\frac{1}{c}{\bf A}({\bf
r})\right]^2 + V({\bf r}).
\end{equation}
The question that we aim to answer in this Letter is: what is the PS
Hamiltonian that, in presence of a magnetic field, satisfies the
requirements (i) and (ii) stated above? The PS Hamiltonian can be written as:
\begin{equation}
H^{\rm PS}_{\bf A} = \frac{1}{2}\left[{\bf p}+\frac{1}{c}{\bf A}({\bf
r})\right]^2 + V^{\rm l}({\bf r}) + \sum_{\bf R} V^{\rm nl}_{\bf R}
+ \Delta H_{\bf A},
\end{equation}
where $\Delta H_{\bf A}$ remains to be determined. $\Delta H_{\bf A}$
cannot be zero for all ${\bf A}({\bf r})$, as $H^{\rm PS}_{\bf A}$ must be
gauge invariant to satisfy (i) and (ii). Its eigenvalues must not
depend on the arbitrary choice of the gauge of ${\bf A}({\bf
r})$. However, as we will demonstrate, demanding gauge invariance
alone is not sufficient to uniquely determine $\Delta H_{\bf A}$.

To obtain $H^{\rm PS}_{\bf A}$ we use Bl\"ochl's projector augmented
wave (PAW) theory \cite{blochl94}.  In the PAW approach the
pseudisation procedure is defined as a linear transformation between
Hilbert spaces -- those of the AE valence wavefunctions, and the
computationally convenient PS wavefunctions. This transformation,
$\ket{\Psi} = {\cal T}\ket{\tilde\Psi}$, can be applied to obtain 
PS operators, $\tilde {\cal O}$, which correspond to their AE
counterparts, ${\cal O}$:
\begin{eqnarray}
\label{operator}
\tilde {\cal O} &=& {\cal T}^\dagger {\cal O} {\cal T} =  {\cal O} + \tilde {\cal C}\\
\label{augmetation}
\tilde {\cal C}&=&
\sum_{i,j}\ket{\tilde
p_i}\left[\bracket{\phi_i}{{\cal O}}{\phi_j} -
\bracket{\tilde\phi_i}{{\cal O}}{\tilde\phi_j}\right]\bra{\tilde p_j},
\end{eqnarray}
where we have adopted Bl\"ochl's tilde to denote a PS quantity. By
construction, the expectation values of $\tilde {\cal O}$ between PS
wavefunctions are equal those of ${\cal O}$ between the corresponding
AE wavefunctions.  The $\ket{\phi_i}$ and $\ket{\tilde\phi_i}$ are
atomic all electron and pseudo partial waves respectively. The
projector functions $\ket{\tilde p_i}$ act only within some augmentation region
-- or core radius in the language of pseudopotential theory. The AE
and PS partial wave coincide outside this augmentation region, and
$\braket{\tilde p_i}{\tilde\phi_i}=\delta_{ij}$.  If the norms of the
AE and PS partial waves are equal and there is just one partial wave
in each angular momentum channel, which we shall assume in the following, then
taking Eq.~(\ref{operator}) with ${\cal O}=H^{\rm AE}$ and ${\bf
B}({\bf r})={\bf 0}$, we obtain the $H^{\rm PS}$ of norm-conserving
pseudopotentials in the Kleinman-Bylander form \cite{kleinbyl}.

In principle, in order to compute expectation values of physical
observables from PS wavefunctions one should use the PS operator
$\tilde{{\cal O}}$.  In practice, since the AE and PS partial waves
are identical outside the augmentation regions and have the same norm inside, 
the term $\tilde {\cal
C}$ in Eq.~(\ref{operator}) can be neglected for a large class of
operators.  For example, the AE perturbation Hamiltonian describing
the coupling with a uniform electric field ${\bf E}$ is $V^{\rm
AE}_{\bf E}({\bf r})={\bf r}\cdot {\bf E}$.  In norm-conserving PS
calculations of the electric-field response-properties, the $\tilde
{\cal C}$ term is always neglected, with $V^{\rm AE}_{\bf E}({\bf r})$
being used as the approximate PS perturbation potential
\cite{eps_sil:baroni,piezo:degironcoli,pol:kingsmith}.  The $\tilde
{\cal C}$ term can also be neglected for operators for which the
weight is concentrated away from the atoms or are almost constant in
the augmentation regions.  This is the case for operators of the form
$[{\bf c\cdot (r-R) }]^n$ in systems with a single PS atom centered at
${\bf R}$. Here, and in the following, ${\bf c}$ is a constant
vector and $n\ge 0$.  In addition, the $\tilde {\cal C}$ term is
exactly zero for the ${\bf R}$-centered angular-momentum operator
${\bf L}_{\bf R}=({\bf r-R}) \times {\bf p}$, as the partial waves
are eigenstates of $|{\bf L}_{\bf R}|^2$ and of $({\bf L}_{{\bf
R}})_z$.  Following this reasoning the $\tilde {\cal C}$ term is also
negligible for operators of the form ${\bf L}_{\bf R}[{\bf c\cdot (r-R)}]^n$.
In contrast, $\tilde {\cal C}$ can not be neglected for the
kinetic energy operator, $p^2/2$, or for operators that are general
functions of the ${\bf p}$ and ${\bf r}$ operators. Indeed, by
construction, the PS wavefunctions can be expanded on a much smaller
set of Fourier components in the momentum space (where ${\bf p}$ is
diagonal) than the corresponding AE wavefunctions.

In general, because of the presence of the ${\bf p}$ operator, the
$\tilde {\cal C}$ term corresponding to the AE magnetic perturbation
Hamiltonian $H^{\rm AE}_{\bf A}-H^{\rm AE}$,
\begin{equation}
\Delta H^{\rm AE}_{\bf A}=\frac{{\bf p}\cdot{\bf A}({\bf r})+{\bf A}({\bf r})\cdot{\bf p}}{2c} +\frac{A({\bf r})^2}{2c^2},
\end{equation}
can not be neglected, even if ${\bf B}({\bf r})$ and ${\bf A}({\bf r})$
are smooth on the scale of the core augmentation regions.  However, as
we shall show, if ${\bf B}({\bf r})$ is smooth, and we consider a
system with a single PS atom centered at ${\bf R}$, there is a special
vector potential gauge, ${\bf A}'({\bf r})$, in which the $\tilde
{\cal C}$ term can be neglected.  This ${\bf A}'({\bf r})$ potential
can be defined in terms of the Fourier components ${\bf b_G}$ of the
magnetic field,
\begin{equation}
{\bf B}({\bf r})=\sum_{{\bf G}}^{|{\bf G}|<G_{\rm max}}{\bf b}_{\bf G}e^{i{\bf
G}\cdot{\bf r}},
\end{equation}
where ${\bf b_{\bf G}}\cdot{\bf G}=0$, and, if the field is smooth on
the scale of the core radius $r_{\rm c}$, then $r_{\rm c}G_{\rm max}\ll 1$.
In particular,
\begin{equation}
\label{magicgauge}
{\bf A}'({\bf r}) = \sum_{{\bf G}}^{|{\bf G}|<G_{\rm max}}\frac{1}{2}{\bf b_G}\times({\bf r}-{\bf R}) 
f[i{\bf G}\cdot({\bf r}-{\bf R})],
\end{equation}
with $f(x)$ defined by,
\begin{equation}
f(x) = 2\frac{1+xe^x-e^x}{x^2}=1+\frac{2}{3}x+O(x^2).
\end{equation}
Note that if the field is smooth, in the core region $x\ll 1$ and $f(x)$ can be 
expanded in powers of $x$.
Since, for a uniform magnetic field ${\bf
A}'({\bf r})$ reduces to the symmetric gauge, ${\bf
A}'({\bf r})$ can be seen as a generalization of the symmetric gauge to
non-uniform magnetic fields.

The magnetic coupling Hamiltonian in the ${\bf A}'({\bf r})$ gauge is:
\begin{equation}
\Delta H^{\rm AE}_{{\bf A}'}=\sum_{{\bf G}}^{|{\bf G}|<G_{\rm max}}
\frac{{\bf b_G}\cdot {\bf L_R}f[i{\bf G}\cdot({\bf r}-{\bf R})]+ cc
}{4c}
+\frac{A'({\bf r})^2}{2c^2},
\end{equation}
where $cc$ stands for the complex conjugate. All the terms in the
asymptotic expansion of ${\bf L_R}f[x]$ are of the form ${\bf
L_R}[{\bf c\cdot(r-R)}]^n$ with $n\ge 0$. As a result, if ${\bf B}({\bf r})$ is
smooth on the scale of the augmentation region, then $\tilde{\cal C}$
term arising from the $\Delta H^{\rm AE}_{{\bf A}'}$ operator can be
neglected. Thus the total PS Hamiltonian in the special ${\bf A}'({\bf
r})$ gauge is:
\begin{equation}
\label{Hmagic}
H^{\rm PS}_{{\bf A}'} = \frac{1}{2}\left[{\bf p}+\frac{1}{c}{\bf A}'({\bf
r})\right]^2 + V^{\rm l}({\bf r}) + V^{\rm nl}_{\bf R}.
\end{equation}

From this result we can show that the Hamiltonian for an arbitrary gauge ${\bf A}({\bf r})$ is:
\begin{eqnarray}
\label{Hsingle}
H^{\rm PS}_{{\bf A}} &=& \frac{1}{2}\left[{\bf p}+\frac{1}{c}{\bf A}({\bf
r})\right]^2 + V^{\rm l}({\bf r}) + \int d^3r'd^3r'' |{\bf r}'\rangle \langle {\bf r}''|\times\nonumber \\
& &V^{\rm nl}_{\bf R}({\bf r}',{\bf
r}'')e^{\frac{i}{c}\int_{{\bf r}'\rightarrow{\bf R}\rightarrow{\bf
r}''}{d{\bf r}\cdot{\bf A}({\bf r})}},
\end{eqnarray}
where ${\bf r}\rightarrow {\bf r}'$ indicates a straight line path
from point ${\bf r}$ to point ${\bf r}'$. To prove Eq.~(\ref{Hsingle}) we have just to notice that $H^{\rm PS}_{{\bf A}}$ is
gauge invariant and reduces to Eq.~(\ref{Hmagic}) for ${\bf A}({\bf r})={\bf A}'({\bf r})$.

Finally, the linearity of the Hamiltonian with respect to the
electron-ion potential can be exploited to obtain $H^{\rm PS}_{{\bf
A}}$ when there is more the one PS atom:
\begin{eqnarray}
\label{Hmulti}
H^{\rm PS}_{{\bf A}} &=& \frac{1}{2}\left[{\bf p}+\frac{1}{c}{\bf A}({\bf
r})\right]^2 + V^{\rm l}({\bf r}) +  \int d^3r'd^3r''|{\bf r}'\rangle \langle {\bf r}''|\times\nonumber \\
& &\sum_{\bf R} V^{\rm nl}_{\bf R}({\bf r}',{\bf
r}'')e^{\frac{i}{c}\int_{{\bf r}'\rightarrow{\bf R}\rightarrow{\bf
r}''}d{\bf r}\cdot{\bf A}({\bf r})}.
\end{eqnarray}
We refer to $H^{\rm PS}_{{\bf A}}$ as the GIPAW Hamiltonian for an arbitrary
magnetic field, since $H^{\rm PS}_{{\bf A}}$ reduces to the Hamiltonian
that we have derived in our earlier work
\cite{pickard2001-1}, if the magnetic field is uniform and if the
symmetric gauge is used.
Our new Hamiltonian holds if the
magnetic field varies smoothly over the core region, and if the
potentials are norm conserving. If the field varies more rapidly, or
the norm conservation is relaxed, the $\tilde{\cal C}$ terms must be
included in the Hamiltonian.

%%%%%%%%%%%%%%%%%%%%%%%%%%%%%%%%%%%%%%%%
%\section*{The magnetic susceptibility}%
%%%%%%%%%%%%%%%%%%%%%%%%%%%%%%%%%%%%%%%%

Our result differs from that of ICL. The Hamiltonian derived 
in Ref.~\cite{ismail-beigi2001} can be obtained from Eq.~(\ref{Hmulti}) if one 
replaces our dog-leg path ${\bf r}'\rightarrow{\bf R}\rightarrow{\bf r}''$ with a straight line path ${\bf
r}'\rightarrow{\bf r}''$. The ICL Hamiltonian is also gauge invariant. In the presence of a magnetic field 
the
value of the integral depends on the path, since
$\nabla\times {\bf A}({\bf r})\ne {\bf 0}$. 
The two gauge invariant Hamiltonians are therefore different.

To clarify the situation, we examine the
consequences of these differences for the calculation of the magnetic
susceptibility.  
To compare the results obtained within the two methods with all-electron calculations,
we restrict ourself to molecular systems for which it is possible
to compute the susceptibility with quantum chemical approaches.
The macroscopic magnetic susceptibility tensor ${\bf \tensor
\chi}$ is defined from the second derivative of the system energy with
respect to the external uniform magnetic field ${\bf B}$:
\begin{equation}
{\bf B}\cdot{\bf \tensor\chi}\cdot{\bf B}=-2E^{(2)}
\end{equation}
where $E^{(2)}$ is the second order variation of the energy with
respect to the magnetic field:
\begin{eqnarray}
\label{E2}
E^{(2)}&=&2\sum_o[\bracket{\tilde\Psi_o^{(0)}}{\tilde H^{(1)}{\cal
G}(\epsilon_o)\tilde H^{(1)}}{\tilde\Psi_o^{(0)}}\nonumber\\
&&+\bracket{\tilde\Psi_o^{(0)}}{\tilde H^{(2)}}{\tilde\Psi_o^{(0)}}],
\end{eqnarray}
$|\tilde\Psi_i^{(0)}\rangle$ and $\epsilon_i$ are the unperturbed eigenstates and eigenvalues,  
${\cal G}(\epsilon_o)=\sum_{e} |\tilde\Psi_e^{(0)}\rangle\langle\tilde \Psi_e^{(0)}|/(\epsilon_o-\epsilon_e)$ and the $o$ and $e$ sums run over occupied and empty orbitals.

In a
uniform magnetic field, with the gauge ${\bf A}({\bf r})={\bf B}\times {\bf r}/2$, our GIPAW Hamiltonian, Eq.~(\ref{Hmulti}), gives
rise to the following perturbation Hamiltonians:
\begin{equation}
\tilde H^{(1)}_{\rm GIPAW}=\frac{1}{2c}\left({\bf L}+\sum_{\bf R}{\bf R}\times{\bf
v}_{\bf R}^{\rm nl}\right)\cdot{\bf B},
\end{equation}
and
\begin{equation}
\tilde H^{(2)}_{\rm GIPAW}=\frac{1}{8c^2}\left[({\bf B}\times{\bf r})^2+\sum_{\bf
R}{\bf B}\times{\bf R}\cdot{\tensor {\bf D}}_{\bf R}^{\rm nl}\cdot{\bf B}\times{\bf
R}\right],
\end{equation}
where $ {\bf v}_{\bf R}^{\rm nl} = [{\bf r},V_{\bf R}^{\rm nl}]/i$,
${D}_{\bf R,\alpha,\beta}^{\rm nl} = -[{r}_\alpha,[{
r}_\beta,V_{\bf R}^{\rm nl}]]$, and $\alpha$ and $\beta$ are Cartesian indexes.

The
corresponding perturbation Hamiltonians obtained following the ICL approach are
\begin{equation}
\tilde H_{\rm ICL}^{(1)}=\frac{1}{2c}\left({\bf r}\times{\bf
v}\right)\cdot{\bf B},
\end{equation}
\begin{equation}
\tilde H_{\rm ICL}^{(2)}=\frac{1}{8c^2}\left[({\bf B}\times{\bf
r})^2+\sum_{\bf R}{\bf B}\times{\bf r}\cdot{\tensor {\bf D}}_{\bf R}^{\rm nl}\cdot{\bf
B}\times{\bf r}\right],
\end{equation}
where ${\bf v} = {\bf p}+\sum_{\bf R}{\bf v}_{\bf R}^{\rm nl}$.

We compute $\tensor \chi$ in molecules with both the GIPAW and the ICL approaches.
We describe the electronic structure with density functional theory in the 
local density approximation.
We use a large-cubic-periodic supercell of 6000 au$^3$, in order to avoid
the interaction of the molecules with their periodic replica, and
Troullier-Martins pseudopotentials~\cite{tm:vps} in the Kleinman-Bylander  form~\cite{kleinbyl}.
We expand the wavefunctions in plane-waves
with a  cutoff of 100 Ry.
The position operator is not defined within
periodic boundary conditions. We treat it approximately by
constructing a periodic saw-tooth like function centered on the molecules.
For large cells this operator well
approximates the position operator where the electron density is not 
negligible~\cite{pickard2001-1}. 
The contribution of core electrons to magnetic properties can not be neglected.
This contribution is however rigid, i.e. independent of the chemical
environment~\cite{gregor99,pickard2001-1}. 
We compute the core contribution
with an atomic code.

\begin{table}
\caption{Gauge invariance test. The magnetic susceptibility of valence electrons
(in cm$^3/$mole) of CH$_4$ is calculated using the GIPAW and ICL
approaches. 
$Tr({\bf \tensor \chi})/3$ is reported as a function of the distance $d$ (in a.u.) of the carbon
nucleus from the gauge origin. The results are decomposed in terms of
the 2 contributions present in Eq.~(\ref{E2}).}
\label{gi}
\begin{tabular}{lrrrrrr}
d   & $\chi_{\rm GIPAW}^{H^{(2)}}$&$\chi_{\rm GIPAW}^{H^{(1)}{\cal G}H^{(1)}}$ &$\chi_{\rm GIPAW}$ &$\chi_{\rm
ICL}^{H^{(2)}}$&$\chi_{\rm ICL}^{H^{(1)}{\cal G}H^{(1)}}$ &$\chi_{\rm
ICL}$ \\\hline
0.0 &-28.4 &8.6 &-19.8 &-28.4 &8.4 &-20.0\\
2.5& -68.0 &48.2& -19.8 &-68.0& 48.0& -20.0\\
5.0 &-186.8 &167.1 &-19.8 &-186.8& 166.9& -20.0\\
7.5 &-384.9 &365.2 &-19.8 &-384.9 &364.9& -20.0\\
10.0& -662.2& 642.5& -19.8 &-662.2 &642.3& -20.0
\end{tabular}
\end{table}

Both the GIPAW and the ICL approaches are, by construction, gauge invariant. 
To verify that our numerical implementation and the use of a finite basis set preserve this property, we compute $\chi$ for a CH$_4$ molecule as a 
function of the distance between the gauge origin and the molecular center.
The results summarized in Table~\ref{gi} show that the calculated $\chi$ is
indeed gauge invariant, while the individual terms
due to ${\tilde H^{(1)}}$ and ${\tilde H^{(2)}}$ clearly are not. 

\begin{table}
\caption{The principal values of the magnetic susceptibility tensor (in cm$^3/$mole) calculated for a
selection of small molecules. The IGAIM results are obtained from the
Gaussian98~\cite{gaussian98} code using the aug-cc-pVQZ basis set for H
and the aug-cc-pCVQZ basis sets for the remaining
elements~\cite{peterson2002}. 
We include in the GIPAW and ICL results the rigid core contributions obtained 
using an atomic code.
The GIPAW and ICL results are
indicated as deviations from the IGAIM results. The root mean square (RMS) and maximum absolute
deviations and percentage deviations for the two methods are reported
at the foot of the table.}
\label{molecules}
\begin{tabular}{lrrrrrrrrrrr}
Mol.      &\multicolumn{3}{c}{IGAIM}  &\phantom{.}& \multicolumn{3}{c}{$\Delta$GIPAW} &\phantom{.}& \multicolumn{3}{c}{$\Delta$ICL}    \\\hline
C$_6$H$_6$    & -99.26 & -33.22 & -33.22& &  -.77 &  -.15 &  -.15 & &  -2.4 & -5.3 & -5.3 \\
CF$_4$        & -33.01 & -33.01 & -33.01& &  -.45 &  -.45 &  -.45 & &  -1.7 & -1.7 & -1.7 \\ 
CH$_3$F       & -25.22 & -16.00 & -16.00& &  -.14 &  -.14 &  -.14 & &  -0.4 & -0.6 & -0.6 \\
CNH$_5$  & -33.59 & -22.83 & -21.71& &  -.12 &  -.16 &  -.16 & &  -0.4 & -0.7 & -0.6 \\
CH$_4$        & -19.87 & -19.87 & -19.87& &  -.06 &  -.06 &  -.06 & &  -0.3 & -0.3 & -0.3 \\
CO            & -18.35 &  -9.63 &  -9.63& &  -.02 &  -.14 &  -.14 & &  -0.1 & -2.2 & -2.2 \\
HCP           & -35.23 & -25.62 & -25.62& &  -.01 &  -.09 &  -.09 & &   0.4 & -2.5 & -2.5 \\
P$_2$         & -46.13 & -18.23 & -18.23& &  -.01 &   .23 &   .23 & &   0.6 & -5.9 & -5.9 \\
PF$_3$        & -28.08 & -27.17 & -27.11& &  -.40 &  -.29 &  -.29 & &  -2.7 & -2.9 & -2.9 \\
Si$_2$H$_4$   & -44.14 & -39.24 & -37.06& &  -.10 &  -.09 &  -.22 & &  -1.5 & -1.6 & -2.9 \\
SiF$_4$       & -40.92 & -40.92 & -40.92& &  -.34 &  -.34 &  -.34 & &  -1.6 & -1.6 & -1.6 \\
SiH$_3$F      & -27.77 & -18.63 & -18.63& &  -.13 &  -.17 &  -.17 & &  -0.9 & -1.2 & -1.2 \\
SiH$_4$       & -23.97 & -23.97 & -23.97& &  -.08 &  -.08 &  -.08 & &  -0.7 & -0.7 & -0.7 \\
              &       &       &      & &       &       &      & &       &       &       \\
RMS $\Delta$  &       &       &      & &\multicolumn{3}{c}{0.25}& &\multicolumn{3}{c}{2.35}\\
Max $|\Delta|$  &       &       &      & &\multicolumn{3}{c}{0.77}& &\multicolumn{3}{c}{5.85}\\  
RMS \%        &       &       &      & &\multicolumn{3}{c}{0.9}& &\multicolumn{3}{c}{11.0}\\
Max $|\% |$        &       &       &      & &\multicolumn{3}{c}{1.9}& &\multicolumn{3}{c}{32.1}  
\end{tabular}
\end{table}

In order to obtain an independent assessment of the accuracy of the two
methods, we compare to a truly all electron method, the individual gauges for atoms in molecules (IGAIM)
method~\cite{nmr:igaim} as implemented in
Gaussian98~\cite{gaussian98}. The magnetic susceptibility converges
only slowly with Gaussian or atomic basis sets. 
The aug-cc-p(C)VxZ basis set
series~\cite{peterson2002} has been previously shown exhibit reliable
convergence for magnetic response
properties~\cite{gregor99,helgaker1999}, and we confirm this by
converging the magnetic susceptibility for CH$_4$ by using up to the
aug-cc-pCV5Z basis for C (and aug-cc-pV5Z for H). For the remaining
calculations we use the corresponding quadruple zeta basis sets, at
which level the CH$_4$ result is converged to better than 0.1
cm$^3/$mole. Indeed,
calculations using the largest basis sets
rapidly become intractable for even moderately sized molecules. The
results are summarized in Table~\ref{molecules}, and show that the
GIPAW method results in values for the magnetic susceptibility that
are consistently closer to the all electron results than those
calculated using the ICL method by roughly an order of
magnitude. E.g., the ICL results for P$_2$ deviate by almost one
third of the total, while the maximum fractional deviation for our
method is less than two percent.

%%%%%%%%%%%%%%%%%%%%%%%
%\section*{Conclusion}%
%%%%%%%%%%%%%%%%%%%%%%%

We have derived, and demonstrated the practical utility of, a theory
for the coupling of nonlocal pseudopotentials to arbitrary
electromagnetic fields. While the ICL method may be the best possible
for an arbitrary nonlocal Hamiltonian we have shown that by focusing on
the nonlocal Hamiltonian most frequently encountered in electronic
structure calculations we are able to make considerable
improvements. This is of great importance if the results are to be
compared quantitatively with experiment.

%%%%%%%%%%%%%%%%%%%%%%%%%%%%%
%\section*{Acknowledgements}%
%%%%%%%%%%%%%%%%%%%%%%%%%%%%%

The work of CJP was supported by the Universit\'es Paris 6 and 7 in
France, and the EPSRC in the United Kingdom. We thank the High
Performance Computing Facility of the University of Cambridge for
access to Hodgkin, their SGI Origin 2000, and the CNRS funded IDRIS
supercomputing center. We would also like to thank Thomas Gregor for
the molecular geometries, and Kirk Peterson for his kind help with the
Gaussian basis functions.

% REFERENCES

%\bibliography{Total} 
%\bibliography{Total} 
%\bibliography{../Total} 

\end{document}